\documentclass[conference]{IEEEtran}
\IEEEoverridecommandlockouts
\usepackage{amsmath, amssymb, amsfonts, graphicx, algorithm, algorithmic, enumerate,color}
\usepackage{cite}
\usepackage{graphicx}
\usepackage{textcomp}

\usepackage{tikz}
\newcommand{\comment}[1]{}

\DeclareMathOperator{\E}{\mathbb{E}}
\def\BibTeX{{\rm B\kern-.05em{\sc i\kern-.025em b}\kern-.08em
    T\kern-.1667em\lower.7ex\hbox{E}\kern-.125emX}}

\newlength\myindent
\newcommand{\squeezeup}{\vspace{-2.5mm}}

\begin{document}
\setlength{\abovedisplayskip}{0pt}
\setlength{\belowdisplayskip}{0pt}

\title{Gridded UAV Swarm for Secrecy Rate Maximization with Unknown Eavesdropper \\
\thanks{This work was supported in part by the EPSRC Project EP/P009670/1, Petroleum Technology Development Fund and University of Nigeria Nsukka.}}

\author{\IEEEauthorblockN{Christantus O. Nnamani\IEEEauthorrefmark{1},
        Muhammad R. A. Khandaker\IEEEauthorrefmark{2} and
        Mathini Sellathurai\IEEEauthorrefmark{3}}
\IEEEauthorblockA{
        \textit{School of Engineering and Physical Sciences, Heriot-Watt University,} \\
        Edinburgh EH14 4AS, United Kingdom \\
        Email: \{\IEEEauthorrefmark{1}con1, \IEEEauthorrefmark{2}m.khandaker, \IEEEauthorrefmark{3}m.sellathurai\}@hw.ac.uk}
}
\maketitle
\begin{abstract} \label{abs}
This paper considers grid formation of an unmanned aerial vehicle (UAV) swarm for maximizing the secrecy rate in the presence of an unknown eavesdropper. In particular, the UAV swarm performs coordinated beamforming onto the null space of the legitimate channel to jam the eavesdropper located at an unknown location. By nulling the channel between the legitimate receiver and the UAV swarm, we obtain an optimal trajectory and jamming power allocation for each UAV enabling wideband single ray beamforming to improve the secrecy rate. Results obtained demonstrate the effectiveness of the proposed UAV-aided jamming scheme as well as the optimal number of UAVs in the swarm necessary to observe a saturation effect in the secrecy rate. We also show the optimal radius of the unknown but constrained location of the eavesdropper.
\end{abstract}


\section{Introduction} \label{intro}
Due to aerial visibility, ease of maneuvering and cost effectiveness, the unmanned aerial vehicles (UAVs) is rapidly becoming a preferred choice for on-demand wireless communication applications \cite{jrnl_6g, Akyildiz_2020}. The benefits accruing to the use of a single UAV as observed in several wireless communication applications can be extrapolated with the use of multiple UAVs, popularly termed as a UAV swarm \cite{obinna_2020, no_fly, Wang_2019}. The control mechanism of the UAV swarm has been discussed in \cite{Intelligence1999, Yeh_2017, Ibrahim2008} for different applications in order to characterize the trade-off between design complexity and performance. In this paper, we explore this trade-off for the deployment of the UAV swarm for physical layer security (PLS).

PLS uses the dynamic intrinsic properties of wireless communication channels to support the legitimate receiver's signal while reducing the information content received by the eavesdropper(s) \cite{obinna_2020}. Such intrinsic channel properties include fading, interference, multi-path, shadowing and noise. Although PLS can be traced to Shannon's theorem \cite{Shannon_1949}, PLS optimization methods have been recently broadened into the use of UAVs \cite{uav_cooperative_jamming, securingUAV_com, secure_uav_uav,uav_secured_com_JTTP}. 
Several use-cases can be found in the literature deploying UAVs for PLS. The use of a single UAV as a security enabler and as a relay node have been explored in \cite{obinna_2020, iterative_uav, mobile_jammer}; delivering artificial noise interwoven with the relayed information signal. It has also been used to deliver classified message to legitimate ground stations amidst eavesdroppers with restrictions of no-fly regions as discussed in \cite{joint_trajectory}. The results obtained in these works showed that the combination of the deployment of UAV jamming and transmission power optimization improves the secrecy performance of the respective models. However, a major bottleneck of these models is the assumption that perfect channel state information (CSI) of the eavesdropper is known to the transmitter. This is impractical especially for passive eavesdroppers that do not transmit or register their locations.

To alleviate this challenge, \cite{robust_trajectory_power} developed algorithms using a UAV as the information source and optimized its trajectory and transmitting power constrained by sparse eavesdroppers within an independent small uncertainty region. Nevertheless, the performance achieved cannot be guaranteed as the uncertainty region expands and overlaps with the certainty region of the legitimate receiver. Similarly, \cite{secure_uav_uav} discussed PLS for a UAV network where the eavesdropper is mobile within the network. In addition to the weak obscurity of the eavesdropper as tracking is enabled, the advantage of aerial visibility of the UAV was undermined.
In continuation to \cite{secure_uav_uav}, investigation of UAV-aided jamming technique for enabling PLS in ground station scenarios where the exact eavesdropper location is unknown was pursued in \cite{obinna_2020}. The unknown eavesdropper location was assumed to be within an ellipse characterizing the coverage region of the transmitter. 
Although this work guarantees positive secrecy rate for the unknown eavesdropper CSI, the secrecy rates reported were relatively low. 

Following positive results from single UAV use-case scenarios for PLS, explorations of techniques to increase the secrecy rate led to the deployment of multiple UAVs in form of a swarm for PLS. The UAV swarm is simply a collection of independent flying UAVs that are autonomous but interact reactively to produce an aggregated behaviour \cite{Intelligence1999, Ibrahim2008}. The theoretical framework for the relationship between the UAV swarm and the ground base station was proffered in \cite{Yin_2019}. In \cite{Wang_2019}, the UAV swarm tracked the movement of the eavesdropper to jam its received signal while maximizing the secrecy rate of the main receivers. The location of the eavesdropper is assumed to be known in this work and the geometry of the UAV swarm was not considered contrary to the specifications suggested in \cite{Yeh_2017, Yin_2019}. However, without considering the geometry of the UAV swarm, the optimized beamforming weights are not guaranteed to produce a beam pattern \cite{Yin_2019}. Therefore, in this paper, we consider a grid formation for the UAV swarm with unknown location of the eavesdropper in order to increase the secrecy rate via coordinated jamming. Simulation results demonstrate the superior performance of the proposed grid-structured UAV swarm approach compared to conventional approaches. 
The technical contribution in this paper can be characterized in terms of seeking the answers to the following questions with respect to existing techniques:
\begin{enumerate}
    \item What is the position of the $k$-th member of the UAV swarm at a particular time instance?
    \item How to harness the properties of the UAV swarm to achieve maximum secrecy rates?
\end{enumerate}
These questions as enumerated are correlated and are answered within the general principles of controlling the UAV swarm and beamforming design proposed in this paper.

\section{System Model and Problem formulation} \label{sec2}
\begin{figure}[!ht]
\centering
\includegraphics[width=\linewidth,height=150cm,keepaspectratio]{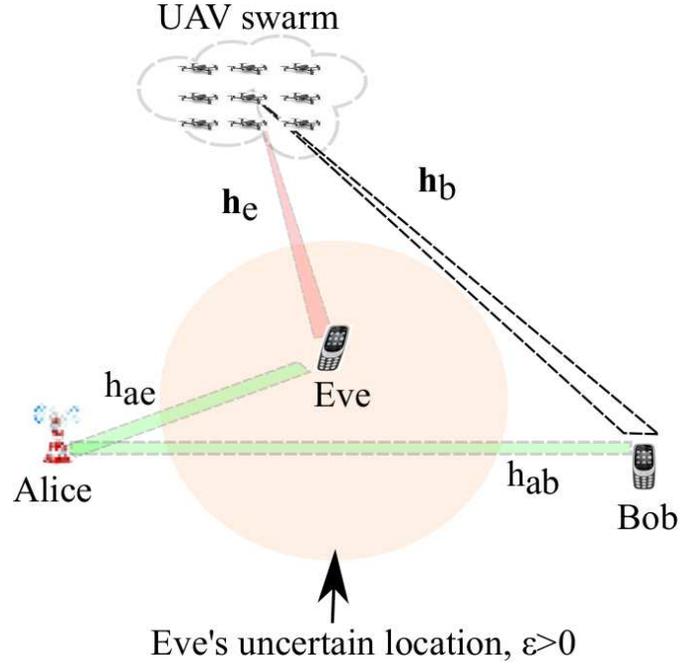}

\caption{UAV Swarm interaction with ground stations}
\label{sys_m}
\end{figure}

We consider a scenario in which a transmitter (Alice) wants to send a confidential message to a legitimate receiver (Bob) in the presence of an eavesdropper (Eve). The primary objective is to ensure that Alice sends the confidential message to Bob without prior knowledge of the presence or location of Eve. A group of $K$ UAVs coordinate their jamming signals to ensure worst channel state of Eve despite its unknown location without tampering the channel of Bob. A pictorial representation is given in Fig.~\ref{sys_m}.

We assume that Eve is located in a closed circular region with radius $\varepsilon$ within the coverage region of Alice. This assumption is feasible given that a passive Eve will practically be within an area where it can easily purloin information without revealing its location. As $\varepsilon \rightarrow 0$, the closer we arrive at the exact location of Eve. However, since the exact position of Eve is unknown, $\varepsilon$ must always be greater than zero and possibly can be increased to cover the entire coverage region of Alice, thereby introducing the maximum uncertainty on Eve's location. Let the location of Alice, Bob, 
and the center of the region of Eve be denoted as $\boldsymbol{\Omega}_{\rm a}$, $\boldsymbol{\Omega}_{\rm b}$, 
${\boldsymbol{\Omega}}_{\rm e}$, respectively. Consider that the entire flight time ($T$) of the UAVs is sampled at discrete time-stamps of $N$ equal time slots, with duration $\delta = \frac{T}{N}$. Without loss of generality, we assume that the UAVs fly at constant altitude, $H$ and with maximum speed of $Z$m/s for each $\delta$ seconds giving rise to a trajectory represented as ${\bf Q}=\{{\bf q}_k[n], n\in N~\&~ k\in K\}$. We note that as $N \rightarrow \infty$, the UAVs are seen as following a continuous trajectory satisfying time-sharing conditions, thereby, delivering jamming signals continuously through its entire flight time \cite{uav_cooperative_jamming}.

The received signal at Bob (b) and Eve (e) in the $n$-th time-slot is given by
\begin{align}\label{rx}
r_i[n]=h_{ai}\underbrace{{w}_{\rm a}s}_{{\rm x}_{\rm a}}+\sum_{k=1}^{K}({h}_{ki}[n]\underbrace{{w}_k[n]\bar {s}[n]}_{{\rm x}_k[n]}) + n_i[n] \\
~{\rm for}~{\rm i}\in \{{\rm b,e}\}, \nonumber
\end{align}
%
%
%
%
where $n_i ~{\rm for}~ i\in\{b,e\}$ is an additive independent and identical white Gaussian noise signal received by Bob or Eve with $ \sim \mathcal{C}(0,\,\sigma_i)\,$, $h_{ai}$ is the complex channel gain between Alice and Bob or Eve while ${\rm x}_{\rm a}$ and ${\rm x}_k$ represents the uncorrelated transmission symbols $(s)$ and jamming signals $(\bar {s})$, respectively. $\{.\}^\ast,~ \{.\}^{\rm T}$ and $\{.\}^{\rm H}$ represent complex conjugate, transpose and Hermitian, respectively at the $n$th sample. We also assume that ${\rm \E}[|{\rm x}_{\rm a}[n]|^2] = \sigma_{\rm a}^2 = 1$. Having prior knowledge of the channel of Bob, we design the jamming signal such that Bob remains in a region free from it. This is achieved by nulling the jamming signal on the channel of Bob for each $n \in \{1,\dots,N\}$, such that the beamforming coefficient (${\rm w}_k$) for each UAV is chosen as to ensure that ${\bf h}_{\rm b}^{{\rm T}}{\bf w} =0$, where ${\bf h}_{\rm b}=[{h}_{1{\rm b}}, ~\dots,~ {h}_{K{\rm b}}]^{{\rm T}}$ and ${\bf w}=[{w}_1, ~\dots,{w}_K]^{{\rm T}}$. This ensures that  ${\bf w}$ lies in the null space of ${\bf h}_{\rm b}$. 
Nevertheless, since the UAV swarm will continue to jam Eve at every possible location within the described uncertainty region, it will imply that ${\bf h}_{\rm e}^{{\rm T}}{\bf w} \neq 0$ for ${\bf h}_{\rm e}=[{h}_{1{\rm e}}, ~\dots,~ {h}_{K{\rm e}}]^{{\rm T}}$.

Furthermore, we define the channel impulse response between the $k$th UAV and the ground stations, ${\rm i}\in \{{\rm b,e}\}$ as \cite{Yin_2019}
\begin{align}\label{hki}
h_{ki}=c_ke^{j\phi_{k}}\delta(t-\tau_k),
\end{align}
where $\phi_k=(\omega_{\rm c}+\omega_{\rm d})t-\omega_{\rm d}\tau_k$ is the phase shift due to Doppler effect $\omega_d$ and time delay $\tau_k;$ $c_k$ is the large scale channel effect due to path loss and shadowing. We assume that there is a line of sight (LoS) communication link between the UAV swarm and the ground stations (Bob and Eve), hence $c_k[n]=\rho_0\varsigma\|{\bf q}_k[n]-\boldsymbol{\Omega}_i\|^{-2}$, where $\rho_0$ represents the channel power gain at reference distance $d_0=1$ m and $\varsigma$ is an exponential random variable with unit mean \cite{obinna_2020, uav_cooperative_jamming, uav_secured_com_JTTP}. 

Accordingly, the signal to interference plus noise ratio $({\gamma})$ at Bob and Eve is given respectively by
\begin{subequations} \label{sinr}
\begin{align}
{\rm {\gamma}}_{\rm b} & =\frac{|h_{\rm ab}|^2}{\sigma_{\rm ab}^2} \label{sinr_b} \\
{\rm {\gamma}}_{\rm e}[n] & = \frac{|{h}_{\rm ae}|^2}{\sum_{k=1}^{K}(|{h}_{ke}[n]{\rm x}_k[n]|^2) + {\sigma}_{\rm e}^2}, \label{sinr_e}
\end{align}
\end{subequations}
where $h_{ai}$ is Rayleigh distributed with gain of ${\rm \E}[|h_{ai}[n]|^2]=\rho_0\varsigma \|\boldsymbol{\Omega}_{\rm a}-\boldsymbol{\Omega}_i\|^{-\mu}$ for ${\rm i}\in \{{\rm b,e}\}$ and $\mu$ is the pathloss \cite{uav_cooperative_jamming}. Since Alice is does not adjust its transmission rate because it is ignorant of Eve and the UAV swarm transmits in the null space of Bob, the channel between Alice and Bob is unaffected by the samples of the UAV as reflected in \eqref{sinr}.

The average secrecy rate is defined as the difference in the information rate of Bob and Eve is given in \eqref{rs} \cite{on_the_secrecy,uav_cooperative_jamming}.
\begin{equation} \label{rs}
R_s = \frac{1}{N}\sum_{n=1}^N [\log_2(1+{\rm {\gamma}}_{\rm b})-\log_2(1+{\rm {\gamma}}_{\rm e}[n])]^+,
\end{equation}
where $[x]^+={\rm max}\{0,x\}$ ensures that the information rate received by Eve is not greater than that received by Bob in a variable rate scheme \cite{on_the_secrecy}. This guarantee of positive secrecy is maintained by setting the power of the transmitter (Alice) when the CSI of Eve is greater than the CSI of Bob. However, if we consider a more realistic scenario where Alice is ignorant of Eve, then it cannot adjust its power based on the CSI of Eve, which may invariably lead to negative secrecy rate if Eve has better CSI than Bob. Hence, the objective of this work is to ensure that the negative secrecy rate is completely mitigated even for cases when Alice is ignorant of Eve. If we assume that the uncorrelated jamming symbols have unity energy, we aim to maximize $R_s$ in \eqref{rs} by finding the appropriate ${\bf w}$
and ${\bf Q}$ which represents the beamforming vectors
and UAV swarm trajectory respectively.

To simplify the trajectory problem, we consider centralized grid swarm control where one element of the swarm is classified as the head of the swarm ${\bf q}_{\rm c}=[x_{\rm c},y_{\rm c},H]$ \cite{Ibrahim2008}. Other elements are distributed in a grid form within a predefined width from the head. This enables the optimization of only the trajectory of the head of the UAV swarm through the optimization process under the constraints in \eqref{cons_q}.
\begin{subequations} \label{cons_q}
\begin{align}
\|{\bf q}_{\rm c}[n+1]-{\bf q}_{\rm c}[n]\|^2\leq(Z\delta)^2 \label{q1} \\
{\bf q}_{\rm c}[N]={\bf q}_f. \label{q2} 
\end{align}
\end{subequations}
Equation \eqref{q1} provides the upper bound for the maximum distance covered by the UAV swarm head within a sample period, while \eqref{q2} constrained its final destination. 
The direct implication of \eqref{q1} ensures that the time of flight of the UAV is lower bound by $T \geq \frac{\|{\bf q}_f-{\bf q}_o\|}{Z}$, where ${\bf q}_f$ and ${\bf q}_o$ represents the final and initial trajectory of the UAV head. This means that the number of discrete time-stamps, $N=\frac{T}{\delta}$, must be sufficient to allow the UAV travel at least in a straight line from its initial to its final point. However, this trajectory is not guaranteed to be optimal in terms of maximizing the secrecy rate. 

The UAV swarm transmit power is bound by the peak and average power constraints given in \eqref{cons_pu} due to the limited capability of each individual UAV payload and power.
\begin{subequations} \label{cons_pu}
\begin{align}
{\rm Tr}({\bf w}[n]{\bf w}[n]^{\rm H}) \leq \bar{P}_{\rm tot}, \label{P_cav_pu} \\
0 \leq |{\rm w}_k|^2 \leq P_{\rm max}, \label{pmax}
\end{align}
\end{subequations}
where $P_{\rm max}$ and $\bar{P}_{\rm tot}$ represents the maximum power transmitted by a single UAV in the swarm and the average power transmitted by the UAV swarm respectively. \eqref{pmax} constrains the minimum and maximum value of the jamming power while \eqref{P_cav_pu} bounds the collective power radiated from the UAV swarm at each $n \in \{1,\dots,N\}$ to minimize external interference. Subsequently, we formulate the UAV swarm problem as
\begin{subequations}\label{p1}
\begin{align}
\max_{{\bf w}, {\bf q}}~& R_s \label{p1a} \\
{\rm s.t.} ~&
{\bf h}_{\rm b}^{T}[n]{\bf w}[n] = 0,\\
&\|{\bf q}_{\rm c}[n+1]-{\bf q}_{\rm c}[n]\|^2\leq(Z\delta)^2,  \\
&{\bf q}_{\rm c}[N]={\bf q}_f, \\ 
& {\rm Tr}({\bf w}[n]{\bf w}[n]^{\rm H}) \leq \bar{P}_{\rm tot}, \\
& 0 \leq |{\rm w}_k[n]|^2 \leq P_{\rm max}.
\end{align}
\end{subequations}
We note that the problem in \eqref{p1} is easily solvable with perfect knowledge of both the channels of Bob and Eve by solving the semi-definite program (SDP) of the successive convex approximation (SCA). Nevertheless, we consider that the location of Eve is unknown within a circular region bound by $\varepsilon$.

\section{Proposed Solution}
Similar to \cite{swipt}, we define the exact location of Eve ($\boldsymbol{\Omega}_{\rm e}$) as a point on a circular uncertain region such that 
\begin{subequations}\label{Eve_loc}
\begin{align} 
& \boldsymbol{\Omega}_{\rm e}=\hat{\boldsymbol{\Omega}}_{\rm e}\pm \Delta \boldsymbol{\Omega}_{\rm e}  \label{Eve_loca}\\
& \|\pm \Delta \boldsymbol{\Omega}_{\rm e}\|=\|\boldsymbol{\Omega}_{\rm e}-\hat{\boldsymbol{\Omega}}_{\rm e}\|\leq \varepsilon, ~{\rm for}~ \varepsilon \geq 0  \\
& \| \Delta \boldsymbol{\Omega}_{\rm e}\| \leq \varepsilon,
\end{align}
\end{subequations}
hold, where $\hat{\boldsymbol{\Omega}}_{\rm e}$, $\Delta \boldsymbol{\Omega}_{\rm e}$ and $\varepsilon$ define the estimated location of Eve, the error of the estimation and the radius of error respectively. 

To solve \eqref{p1}, we decompose the problem into a two (2) sub-problems describing the beamforming vectors and the UAV swarm trajectory. The original problem \eqref{p1} can be solved iteratively between the sub-problems to obtain the sub-optimal/near optimal results that satisfies the constraints as used in \cite{uav_cooperative_jamming, jrnl_secrecy, obinna_2020}.
\subsection{Solving for beamforming vectors (${\bf w}$)}

Decomposing \eqref{p1} into \eqref{pr1} and considering only constraints that relates to beamforming vectors, we have that,
\begin{subequations} \label{pr1}
\begin{align} 
\max_{{{\bf W}}} ~& \sum_{n=1}^N \left[ \log_2 \left(1+\left(\frac{|h_{\rm ab}|^2}{\sigma_{\rm b}^2}\right)\right) \right. \nonumber\\
&\left. -\log_2 \left(1+\left(\frac{\frac{1}{\sigma_{\rm e}^2}|h_{\rm ae}|^2}{\frac{1}{\sigma_{\rm e}^2}\sum_{k=1}^K (|h_{ke}[n]{w}_k[n]|^2)+1}\right)\right) \right] \\
{\rm s.t.}~&
{\bf h}_{\rm b}^{T}[n]{\bf w}[n] = 0,\\
& {\rm Tr}({\bf w}[n]{\bf w}[n]^{\rm H}) \leq \bar{P}_{\rm tot},\\
& 0 \leq |{\rm w}_k[n]|^2 \leq P_{\rm max}. \label{bound}
\end{align}
\end{subequations}

Since the power transmitted by the UAV swarm at each $n$-th sample of the trajectory is independent of other samples, the problem in \eqref{pr1} can be simplified by obtaining the values of $w_k$ for each $n$. Hence, we can rewrite \eqref{pr1} for each $n \in \{1,\dots,N\}$ as
\comment{
Since the power transmitted by the UAV swarm at each $n$-th sample of the trajectory is independent of other samples, the problem in \eqref{p1} can be simplified by obtaining the values of $w_k$ for each $n$. Hence, we can rewrite \eqref{p1} for each $n \in \{1,\dots,N\}$ as \eqref{pr2}, by considering only constraints that relates to beamforming vectors.
}
\begin{subequations} \label{pr2}
\begin{align} 
\max_{{{\bf W}}} ~& \log_2\left(1+\left(\frac{|h_{\rm ab}|^2}{\sigma_{\rm b}^2}\right)\right)  \nonumber \\
&  -\log_2\left(1+\left(\frac{\frac{1}{\sigma_{\rm e}^2}|h_{\rm ae}|^2}{\frac{1}{\sigma_{\rm e}^2}{\bf h}_{\rm e}^{T}{\bf W}{\bf h}_{\rm e}+1}\right)\right), \forall{n}\\
{\rm s.t.}~&
{\bf h}_{\rm b}^{T}{\bf w} = 0 \label{orth}\\
& {\rm Tr}({\bf W})\leq \bar{P}_{\rm tot} \\
& {\rm diag}({\bf W}) \leq P_{\rm max} \label{dia_cons}\\
& {\rm rank}({\bf W})=1, \label{rank1con}
\end{align}
\end{subequations}
where \eqref{dia_cons} is a reformation of \comment{\eqref{pmax}} \eqref{bound} and \eqref{rank1con} is a corollary of ${\bf W}={\bf ww}^{{\rm H}}$. Note that \eqref{orth} is a condition necessary to fulfil the nulling of the channel of the main receiver, Bob. Hence, we desire to find some ${\bf w}$ such that \eqref{orth} will be satisfied. For simplicity, we satisfy this condition by obtaining a set of complex vectors, ${\bf v}$, such that ${\bf w}=\{{\bf v} | {\bf h}_{\rm b}^{{\rm H}}{\bf v}=0\}$. ${\bf v}$ becomes the projection vector onto the subspace of ${\bf w}$. Hence, ${\bf w}={\bf H}_{\bot}{\bf v}$; where ${\bf H}_{\bot}$ is the transformation matrix for the projection of ${\bf v}$ on ${\bf w}$. It is apparent that ${\bf W}={\bf H}_{\bot}{\bf v}({\bf H}_{\bot}{\bf v})^{{\rm H}}={\bf H}_{\bot}{\bf vv}^{{\rm H}}{\bf H}_{\bot}^{{\rm H}}$ (\cite[eq. 1.9]{matrix_cookbook}). Let ${\bf V}={\bf vv}^{{\rm H}}$, using \cite[eq. 1.1.17]{matrix_cookbook}, \eqref{pr2} can be reformulated as
\begin{subequations} \label{pr3}
\begin{align} 
\max_{{{\bf V}}} ~& \log_2\left(1+\left(\frac{|h_{\rm ab}|^2}{\sigma_{\rm b}^2}\right)\right) \nonumber\\
& -\log_2\left(1+\left(\frac{\frac{|h_{\rm ae}|^2}{\sigma_{\rm e}^2}}{\frac{1}{\sigma_{\rm e}^2}{\rm Tr}({\bf h}_{\rm e}{\bf h}_{\rm e}^{T}({\bf H}_{\bot}{\bf VH}^{{\rm H}}_{\bot}))+1}\right)\right), \forall{n}\\
{\rm s.t.}~&
{\rm Tr}({\bf H}_{\bot}{\bf VH}^{{\rm H}}_{\bot})\leq \bar{P}_{\rm tot} \\
& {\rm diag}({\bf H}_{\bot}{\bf VH}^{{\rm H}}_{\bot}) \leq P_{\rm max} \\
& {\rm rank}({\bf V})=1. \label{rank1con2}
\end{align}
\end{subequations}
The problem in \eqref{pr3} is a non-convex SDP problem \cite{jrnl_int}. Hence, using the method for solving an SDP problem obtained in \cite{Hui_Ming_2013}, we omit the rank constraint. We note that due to the grid formation of the swarm, ${\bf V}$ is symmetric and represents the tensors (outer product) of ${\bf v}$ and ${\bf v}^{{\rm H}}$, the ${\rm rank}({\bf V})=1$ is guaranteed provided that the ${\bf v}$ is a non-zero vector \cite{Pierre_2008}. Since Alice is assumed ignorant of Eve and subsequently transmit continuously through the flight of the UAV swarm, the UAV swarm will continually send jamming signal through out the entire flight duration. This ensures that the vector, ${\bf v}$ is not zero for each $n$-th sample. Hence, neglecting the constant terms of \eqref{pr3}, we reformulate as \eqref{pr4}. 
\begin{subequations}\label{pr4}
\begin{align} 
\max_{{{\bf V}}} ~& {\rm Tr}({\bf h}_{\rm e}{\bf h}_{\rm e}^{T}({\bf H}_{\bot}{\bf VH}^{{\rm H}}_{\bot}))+1, ~\forall{n}\\
{\rm s.t.} ~&
{\rm Tr}({\bf H}_{\bot}{\bf VH}^{{\rm H}}_{\bot})\leq \bar{P}_{\rm tot} \\
& {\rm diag}({\bf H}_{\bot}{\bf VH}^{{\rm H}}_{\bot}) \leq P_{\rm max}. 
\end{align}
\end{subequations}
We note that the solution obtained in \eqref{pr4} is sufficient to characterize the suboptimal solution of \eqref{pr3}. \eqref{pr4} is a convex SDP problem that can be efficiently solved using SDPT3 solvers \cite{cvx1}.
\subsection{Solving for trajectory of the UAV swarm $({\bf q})$}
To obtain the trajectory of the UAV swarm, we optimize the trajectory of the head of the swarm and derive the trajectory of other members of the UAV swarm in relation to the head since they are in grid formation with fixed spacing. If we consider the scenario when the location of Eve is unknown but can be estimated to exist within \eqref{Eve_loc}. Problem \eqref{p1} can be reformulated in terms of trajectory of the head UAV swarm as given in \eqref{Q2}. 
\begin{subequations} \label{Q2}
\begin{align} 
\max_{{{\bf Q}_{\rm c}}} ~& \sum_{n=1}^N \left[   \log_2\left(1+\left(\frac{|h_{\rm ab}|^2}{\sigma_{\rm b}^2}\right)\right) \right. \nonumber\\
&\left. -\log_2\left(1+\left(\frac{\gamma_0\|\boldsymbol{\Omega}_{\rm a}-(\hat{\boldsymbol{\Omega}}_{\rm e}-\Delta \boldsymbol{\Omega}_{\rm e})\|^{-\mu}}{\frac{\gamma_0p_{\rm u}[n]}{\|{\bf q}_{\rm c}[n]-(\hat{\boldsymbol{\Omega}}_{\rm e}-\Delta \boldsymbol{\Omega}_{\rm e})\|^2}+1}\right)\right)\right ]\\
{\rm s.t.} ~&
{\rm Constraint}~\eqref{cons_q},
\end{align}
\end{subequations}
where $\gamma_0[n]=\frac{\rho_0}{\sigma_{\rm e}^2[n]}$, and $p_{\rm u}=|{\bf w}_{k}|^2$ represents the signal to interference noise ratio at reference distance $d=1m$ and the transmit power from swarm head. Since the location of Alice and Bob are fixed, we assume that the noise variation is the same for each $n\in N$.
Using triangular inequality for ${\bf x}\in\{{\bf q}_{\rm c}[n],\boldsymbol{\Omega}_{\rm a}\}$, 
 and substituting \eqref{Eve_loc}, we have that
\begin{align} \label{dist_cons}
&\|{\bf x}-(\hat{\boldsymbol{\Omega}}_{\rm e}-\Delta \boldsymbol{\Omega}_{\rm e})\| \leq \|{\bf x}-\hat{\boldsymbol{\Omega}}_{\rm e}\|+\varepsilon. 
\end{align}
The right hand side is a lower bound to euclidean distance between the UAV swarm head and the center of the circular region\footnote{Due to the approximation of \eqref{dist_cons}, $\varepsilon=0$ does not represent the case for when the exact location of Eve is known, however, it gives an insight into the goodness of the estimator. If $\varepsilon \to 0$, then Eve is bound to be located at the center of the region of uncertainty.} in which Eve is located. The lower bound represents the best case scenario since the influence of the jamming signal of the swarm will be greater if it is close to Eve. On the contrary, the lower bound represents the worst case scenario for the transmitter (Alice), since it gives the closest euclidean distance between Alice and Eve. If Eve is close to Alice the likelihood of it to purloin information increases. Thus, \eqref{Q2} can be rewritten with bounds as
\begin{subequations} \label{Q3}
\begin{align} 
\max_{{{\bf Q}_{\rm c}}} ~& \sum_{n=1}^N \left[   \log_2\left(1+\left(\frac{|h_{\rm ab}|^2}{\sigma_{\rm b}^2}\right)\right) \right. \nonumber\\
&\left. -\log_2\left(1+\left(\frac{\gamma_0(\|\boldsymbol{\Omega}_{\rm a}-\hat{\boldsymbol{\Omega}}_{\rm e}\|+\varepsilon)^{-\mu}}{\frac{\gamma_0p_{\rm u}[n]}{(\|{\bf q}_{\rm c}[n]-\hat{\boldsymbol{\Omega}}_{\rm e}\|+\varepsilon)^2}+1}\right)\right) \right] \label{Q3a}\\
{\rm s.t.} ~&
{\rm Constraint}~\eqref{cons_q}.
\end{align}
\end{subequations}
Note that \eqref{Q3} is non-convex due to \eqref{Q3a} but it can be solved by introducing  slack variable, $M=\{m[n]=(\|{\bf q}_{\rm c}[n]-\hat{\boldsymbol{\Omega}}_{\rm e}\|+\varepsilon)^2, ~n \in \{1,\dots,N\}\}$ to characterize the separation between the UAV and Eve.
\begin{subequations} \label{Q1a}
\begin{align} 
\max_{{{\bf Q}_{\rm c}},{\bf M}} ~& \sum_{n=1}^N \left[\log_2\left(1+\left(\frac{|h_{\rm ab}|^2}{\sigma_{\rm b}^2}\right)\right) \right. \nonumber\\
&\left. -\log_2\left(1+\left(\frac{\gamma_0(\|\boldsymbol{\Omega}_{\rm a}-\hat{\boldsymbol{\Omega}}_{\rm e}\|+\varepsilon)^{-\mu}}{\frac{\gamma_0p_{\rm u}[n]}{m[n]}+1}\right)\right) \right]\label{Q1aa}\\
{\rm s.t.} ~&
(\|{\bf q}_{\rm c}[n]-\hat{\boldsymbol{\Omega}}_{\rm e}\|+\varepsilon)^2-m[n] \leq 0, \\
& {\rm Constraint}~\eqref{cons_q}.
\end{align}
\end{subequations}
However, \eqref{Q1a} is till non-convex due to the maximization of a convex function of \eqref{Q1aa}. It can be solved using successive convex approximation (SCA) technique given in \cite{sca}. This allows to solve a local tight approximation under tight constraints and relax until the original problem is solved. Hence, given a predefined initial feasible trajectory, ${\bf Q}_{\rm c}^{\rm l}[n]=\{{\bf q}_{\rm c}^{\rm l}[n], n \in \{1,\dots,N\}\}$ for the $l{\rm -th}$ iteration, the non-constant term of the objective of \eqref{Q1a} can approximated with the first order taylor expansion as given in \eqref{tay}
\begin{align} \label{tay}
\log_2\left(1+\left(\frac{\gamma_0(\|\boldsymbol{\Omega}_{\rm a}-\hat{\boldsymbol{\Omega}}_{\rm e}\|+\varepsilon)^{-\mu}}{\frac{\gamma_0p_{\rm u}[n]}{m[n]}+1}\right)\right) \nonumber \\
\leq F^{\rm l}[n](m[n]-m^{\rm l}[n])+G^{\rm l}[n]
\end{align}
where
\begin{align*}
F^{\rm l}[n]=&[\gamma_0^2(\|\boldsymbol{\Omega}_{\rm a}-\hat{\boldsymbol{\Omega}}_{\rm e}\|+\varepsilon)^{-\mu}p_{\rm u}[n]] \\
& *[(m^{\rm l}[n]+\gamma_0p_{\rm u}[n])((\gamma_0(\|\boldsymbol{\Omega}_{\rm a} \\
& -\hat{\boldsymbol{\Omega}}_{\rm e}\|+\varepsilon)^{-\mu}+1)m^{\rm l}[n]+\gamma_0p_{\rm u}[n])]^{-1},
\end{align*}
\begin{align*}
m^{\rm l}[n]=(\|{\bf q}_{\rm c}^{\rm l}[n]-\hat{\boldsymbol{\Omega}}_{\rm e}\|+\varepsilon)^2, ~~~~~~~~~~~~~~~~~~~~~
\end{align*}
\begin{align*}
G^{\rm l}[n]=\log_2\left(1+\frac{\gamma_0(\|{\boldsymbol{\Omega}}_{\rm a}-\hat{\boldsymbol{\Omega}}_{\rm e}\|+\varepsilon)^{-\mu}m^{\rm l}[n]}{m^{\rm l}[n]+\gamma_0p_{\rm u}[n]}\right).
\end{align*}
Neglecting all constant terms in \eqref{tay}, \eqref{Q1a} can be reformulated as
\begin{subequations}\label{Q1b}
\begin{align} 
\max_{{{\bf Q}_{\rm c}},{{\bf M}}} ~& \log_2\left(1+\left(\frac{|h_{\rm ab}|^2}{\sigma_{\rm b}^2}\right)\right)-F^{\rm l}[n]m[n]\\
{\rm s.t.} ~&
(\|{\bf q}_{\rm c}[n]-\hat{\boldsymbol{\Omega}}_{\rm e}\|+\varepsilon)^2-m[n] \leq 0 \\
& {\rm Constraint}~\eqref{cons_q}.
\end{align}
\end{subequations}
Now, \eqref{Q1b} is convex and can be efficiently solved using interior point method in cvx \cite{cvx1}. Since \eqref{Q1b} maximizes the lower bound of \eqref{Q1a}, the objective value obtained is at least equal to \eqref{Q1a} using the updated trajectory, ${\bf Q}^l[n]$.  In the following section, we evaluate the performance of the UAV swarm with Algorithm~1.

\begin{algorithm} [!ht]\label{algo1}
    \caption{Iterative algorithm for solving ${\bf w},~\textrm{and}~{\bf Q}$}
  \begin{algorithmic}[1]
    \STATE Initialize ${\bf w}~\textrm{and}~ {\bf Q}_{\rm c}$ such that the constraints in \eqref{cons_pu} and \eqref{cons_q} are satisfied.
    \STATE $m \leftarrow 1.$
    \STATE \textbf{repeat}
    \begin{ALC@g}
    \STATE Using the grid parameters, construct the location of all the other $K$ UAVs in the swarm, giving rise to ${\bf Q}.$ 
    \STATE Determine the channel impulse response between the UAV swarm and the ground nodes from \eqref{hki}.
    \STATE Compute and update ${\bf w}$ with ${\bf Q}$ by solving \eqref{pr4}.
    \STATE Compute $R_s$ as defined in \eqref{rs}.
    \STATE $e=\bigg|\frac{R_s^{new}-R_s^{old}}{R_s^{new}}\bigg |$.
    \STATE $m \leftarrow m + 1.$
    \STATE Using updated ${\bf w},$ solve \eqref{Q1b} and update ${\bf Q}_{\rm c}$.
    \end{ALC@g}
    \STATE \textbf{until} {$e< \theta$ OR $m\geq m_{max}$.}
    \STATE \textbf{Output:} ${\bf w}~\textrm{and}~\boldsymbol {\rm Q}$.
  \end{algorithmic}
\end{algorithm}

\section{Results and Analysis}
\begin{figure}[!ht]
\centering
\includegraphics[width=\linewidth]{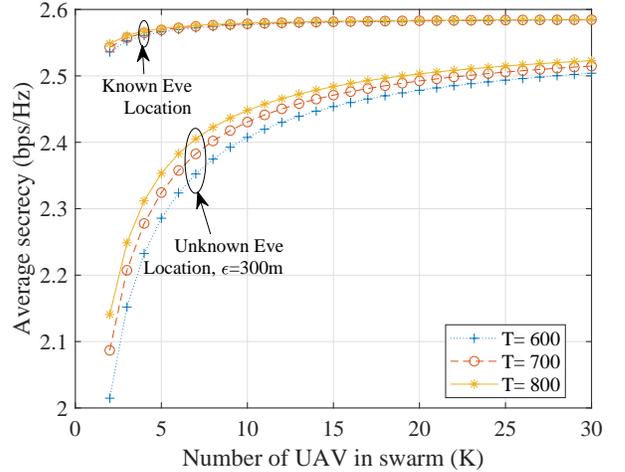}
\squeezeup
\caption{Comparative performance of the UAV Swarm on average secrecy rate when the eavesdropper location is known and unknown.}
\label{asy}
\end{figure}

\begin{figure}[!ht]
\centering
\includegraphics[width=\linewidth]{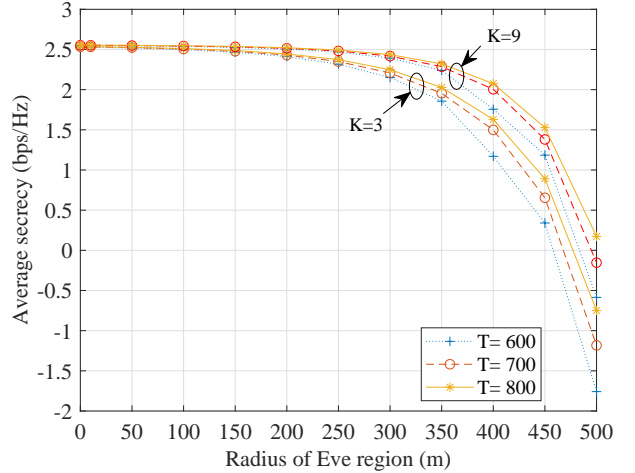}
\squeezeup
\caption{Effect of radius of Eve region on average secrecy.} 
\label{est}
\end{figure}

\begin{figure}[!ht]
\centering
\includegraphics[width=\linewidth]{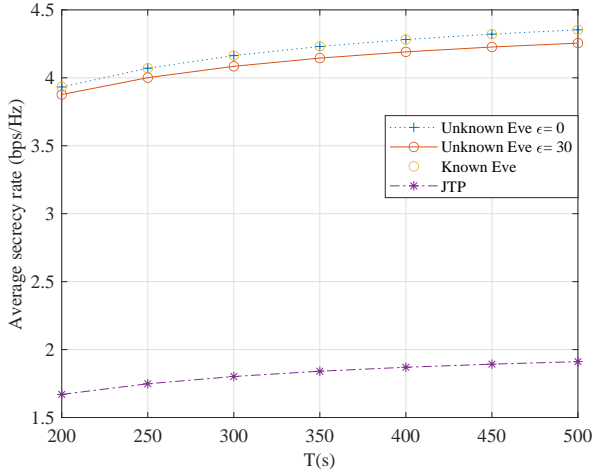}
\squeezeup
\caption{Comparative performance of the average secrecy rate of the UAV Swarm and single UAV jammer for $K=9, ~\boldsymbol{\Omega}_{\rm b}=[200,0,0]^T, ~\hat{\boldsymbol{\Omega}}_{\rm e}=[200,150,0]^T, ~{\bf q}_f=[150,100,100]^T$.  }
\label{tof}
\end{figure}

\begin{figure}[!ht]
\centering
\includegraphics[width=\linewidth]{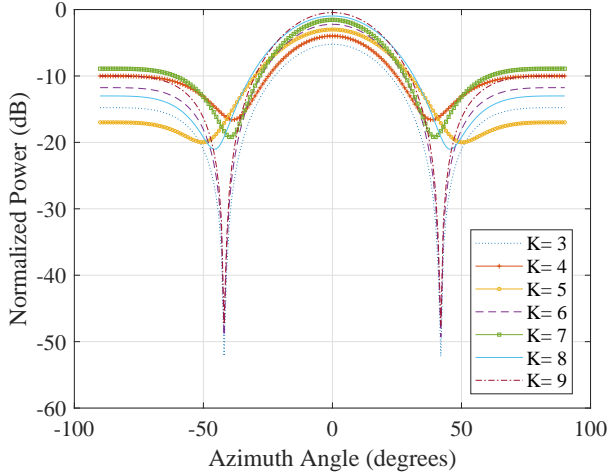}
\squeezeup
\caption{Beam Pattern at $T=600$ and $\varepsilon=300m$.}
\label{rad}
\end{figure}

\begin{figure}[!ht]
\centering
\includegraphics[width=\linewidth]{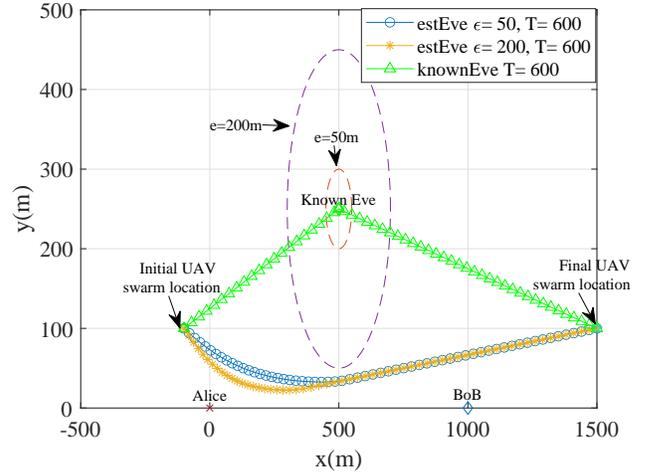}
\caption{The UAV swarm trajectory.}
\label{tra}
\end{figure}

We evaluate the performance of the UAV swarm with parameters set as follows unless otherwise stated: $\boldsymbol{\Omega}_{\rm a}=[0,0,0]^T$, $\boldsymbol{\Omega}_{\rm b}=[1000,0,0]^T,~\hat{\boldsymbol{\Omega}}_{\rm e}=[500,250,0]^T$, $Z=3$m/s, $\gamma_0=90$dB, ${\bf q}_o=[-100,100,100]^T$, ${\bf q}_f=[1500,100,100]^T$, $\delta=1$s, $\bar{P}_{\rm tot}=20$dBm, $P_{\rm max}=26$dBm, $\mu=3.1$ (outdoor), ${\rm Grid~gutter}=10\lambda$, ${\rm Grid~cell}=3$. The initial values of the optimization parameters satisfying respective constraints were obtained via feasibility analysis. The UAV swarm beamforming vectors and its trajectory were solved by iteratively optimizing each parameter with the knowledge of the others, until the error ($e$) between steps is less than $\theta$ (where $\theta=10^{-5}$) or the maximum number of iterations is reached (where $m_{max}=200$). In the legends in Figs.~\ref{asy} - \ref{tra}\comment{\ref{tof}}; $T$, $K$, and $e$ represents the UAV swarm flight time, the number of UAVs and the radius ($\varepsilon$) of the location of Eve respectively. While $estEve$ and $knownEve$ represents the trajectory plots when Eve location is unknown and when it is known respectively.
\squeezeup
\squeezeup
\squeezeup
\squeezeup
\squeezeup
\squeezeup

The performance of the UAV swarm in relation to PLS at different flight times is given in Fig.~\ref{asy}. It shows that as the number of UAVs in the swarm increases, the improvement on the average secrecy rate converges. Hence, considering the power constraints of UAVs, determining the minimum number of UAVs required to attain the secrecy rate convergence is necessary for resource management purposes. However, the effect of $K$ is minimal when the exact position Eve is known. Furthermore, as the number of UAVs making up the swarm increases, the flight time of the swarm does not influence the secrecy rate. 
\squeezeup
\squeezeup
\squeezeup


Further observation on Fig.~\ref{est} presents the limit to the uncertainty location area of Eve at different flight times. As the radius where Eve is located increases, the impact of the UAV swarm jamming signal becomes less significant as the average secrecy rate reduces. Negative average secrecy rate is obtained once Eve has better CSI than Bob and the UAV swarm do not provide enough jamming power as shown in Fig.~\ref{est}. The influence of the UAV swarm flight time is further increased at larger radius. Nevertheless, the characterization of the maximum tolerable error radius in the position of Eve will guide the design of its location estimators.

In Fig.~\ref{tof}, the average secrecy rate performance analysis of the UAV swarm joint trajectory and beamforming optimization for known and unknown Eve's location are compared with known Eve location scenario considered in \cite{uav_cooperative_jamming} (referred to as JTP in the legend) under similar power constraint. It is evident from the figure that the application of the UAV swarm out-performs the baseline scheme of a single UAV jamming model. It can be further observed that the longer time of flight of the UAVs ensures better secrecy performance. This is intuitively, and correlates with similar results presented in \cite{joint_trajectory, uav_cooperative_jamming, securingUAV_com}, since the UAVs delivers more jamming signal during the communication with longer duration of flight time. When the radius of error $\varepsilon$ is zero, the scenario where Eve location is unknown relaxes to known Eve's location since Eve can only be located at a single point. However, as $\varepsilon>0$, the average secrecy rate reduces (as also presented in Fig.~\ref{est}).

Examining the radiation pattern generated by the UAV swarm, as presented in Fig.~\ref{rad}, 
it is observed that for $K$ values that did not complete a quadrilateral formation of the grid $(K=4,5,7,8)$, the null depth is shallow compared to values where the grid quadrilateral is complete $(K=3,6,9)$. The implication is that little spurious jamming signal from the UAV swarm has greater tendencies of leaking to Bob when the grid formation is incomplete. Despite the observation that higher values of $K$ gives higher power in the main lobe, it is apparent that even when multiple UAVs are available, a selection needs to be made to ensure the grid quadrilateral is complete with recourse to the minimum number required to achieve maximum average secrecy rate as shown in Fig.~\ref{asy}. However, for all values of $K$, side lobes with high power levels are observed. Although in conventional beamforming, the aim is to minimize the side lobes, however, since the exact position of Eve is unknown and Bob is at the null of the jamming signal, the jamming power radiated from the side lobes will further reduce the information content received by Eve, especially where multiple Eve exist. 

The trajectory of the swarm in Fig.~\ref{tra}, shows that the optimal trajectory when Eve's location is unknown follows a path close to Alice and Bob. Since the channel models are distance dependent, if Eve enjoys better channel quality than Bob then the zero or negative average secrecy rate ensues. The UAV prioritizes sending higher jamming signal to Eve when it is closer to Alice. Since it cannot determine the exact location of Eve, following this trajectory ensures that the channel for Eve is always degraded despite its proximity to the transmitter (Alice). On the contrary, when the location of Eve is known (knownEve T=600 in Fig.~\ref{tra}), the swarm flies directly above it and jams the signal. 

\section{Conclusion}
We have exploited the idea of grid-formed UAV swarm to demonstrate that the system can improve the secrecy rate of ground communications. Based on the results obtained, we defined the optimal trajectory of the UAV swarm harnessing the optimal number of UAVs in the grid to improve PLS. Furthermore, we examined the impact of the radius of the eavesdropper's constrained location on the secrecy rate and demonstrated the need to develop techniques to increase the directivity of the UAV swarm.

\bibliographystyle{IEEEtran}
\footnotesize{
\bibliography{IEEEabrv, ref_swarm}
}

\end{document}